\documentstyle[11pt]{article}
 3
 2
\def\funfn{\tt}
\def\exfn{\bf}

\def\matfn{\sl}
\def\abvt{\vskip 30 pt}
\def\belt{\vskip 10 pt}
\def\stfn{\bf}
\font\dillfn = cmss12
\def\Dill{{\dillfn Dill}}
\font\bbsf = cmssbx10 scaled\magstep 3

\begin{document}

\hskip 4 in
hep-th/9412298

\vglue 0.5 in
\centerline{{\bbsf Dill:}
{\LARGE\bf
An Algorithm and a Symbolic Software Package
}}

\centerline{\LARGE\bf
for Doing Classical Supersymmetry Calculations\footnote
{\rm The work was supported in part by DOE grant
no. DE-FG02-85ER40233}
}

\vskip 0.7 in
\centerline{
Vladan Lu\u ci\'c
}
\centerline{\it
Physics Department
}
\centerline{\it
Northeastern University, Boston, MA 02115, USA
}
\centerline{\it
e-mail: vlucic@lynx.neu.edu
}

\vskip 1 in
\centerline {\Large\bf Abstract}
\belt

An algorithm is presented that formalizes different steps in a
classical
Supersymmetric (SUSY) calculation. Based on the algorithm \Dill , a
symbolic software package, that can perform the calculations, is
developed in the {\matfn Mathematica} programming
language. While the algorithm is quite general, the package is created
for the $4-D,\ N=1$ model. Nevertheless, with little modification, the
package could be used for other SUSY models. The package
has been tested and some of the results are presented.

\vfill \eject



\section{Introduction}

SUSY and supergravity theories have been candidates for
unifying electro-weak, strong and possibly gravitational interactions
\cite{Mohapatra}.
They are also used as effective theories, describing the low-energy
behavior of a more fundamental theory, as is the case with the string
theories \cite{Cas:strings}.
Many calculations in such theories, on the classical level, include
tedious but conceptually straightforward manipulations. A typical
object that is
manipulated is an expression that contains
both commuting and anticommuting spinor and tensor fields as well as
superfields
with different indices. Typically, one has
space-time, spinor and indices denoting extended representations
(for $N > 1$).
Furthermore, a field may have a definite symmetry with respect to
the interchange of some of its indices. Also, since at least some
of the fields are anticommuting, there are nontrivial symmetries with
respect to interchange of the order of the fields. All this makes it
more difficult to keep track of signs, to recognize and group or
cancel the same terms, or to eliminate the terms that vanish due to
the symmetry properties.
There are also operations that are often used, such as the
application of
different kinds of derivatives: ordinary, Grassman, or covariant. Again,
the rules are clear, but the actual application
can be tedious.

In recent years,
symbolic programming packages have been developed for different
problems, such as calculation of Clebsh-Gordan and Rachah coefficients
\cite{Klink}, tensor products of Lie algebras \cite{Wbase}, in
differential geometry \cite{TTC} \cite{EXCALC}, general
relativity \cite{MacCullum} \cite{Schrufer}, gamma matrices
manipulation \cite{Tracer}, and for the
calculation of Feynman diagrams \cite{HIP}, to mention just a few.
When the importance of SUSY was realized, symbolic
packages were developed for problems related to it:
super-Feynman diagrams
\cite{SYSCAL}, Lie superalgebra manipulations \cite{Cecchini},
calculation of the supertrace of a supermatrix \cite{epicGRASS} and
the anomaly calculation in superstring theories \cite{SUPERCALC}, among
others.
Having in mind the manipulations needed in actual SUSY and
supergravity calculations, it seemed
natural to develop an algorithm  and implement it in one
of the symbolic programming languages that could perform some of the
manipulations,
thereby reducing the effort one has to make in performing the
calculations. The idea is not new; a package was developed in {\matfn
REDUCE} \cite{REDUCE}
for exterior calculus in superspace and the construction of
supergravity actions \cite{Castellani}. In the last few years some very
sophisticated and flexible symbolic programming systems were developed
(e.g. {\matfn Mathematica}
\cite{Mathematica} and {\matfn Maple}
\cite{Maple}) and used in different symbolic programming projects
\cite{TTC} \cite{Tracer} \cite{HIP}.
An algorithm and a package based on a symbolic programming
system, that could perform analytical calculations in SUSY,
would make it possible not only to use the package by itself, but also
to use the
commands of the symbolic system to further process the output of the
package, or to integrate the package with some other, giving the user
more flexibility in using the package for different problems.
Hence, \Dill , the software package that can perform classical SUSY
calculations, has been developed in {\matfn Mathematica 2.0}. It
consists
of several separate modules, or functions each doing a specific
operation on a given expression.
One of the most important functions renames dummy indices and
reorders factors in a product in a unique way,
so that mathematically equal terms have the same form, therefore
making it possible for
the underlying symbolic programming system to group the same terms in
the expression that is manipulated. The functions can be used
separately, or, when a certain sequence of operations is needed,
as commands in a program that does the whole sequence.
In general, a calculation in SUSY involves many different
steps; some of them are technical and straightforward, so they are
included in the package, but some require more inventive, or ``human",
manipulations. Therefore, \Dill\ is
intended to be used interactively; whenever a user who, while doing a
complicated calculation, needs to perform one or more of the
calculations
defined in the package, he/she simply applies the appropriate
function(s), gets a result and continues the calculation on his/her own,
using \Dill\ when needed. In order to use the package, one has to
have a working knowledge of {\matfn Mathematica}.

An application of a function returns an expression in such a format
that another function can be applied on the returned expression.
Unfortunately, it is not easy for a user to read such an expression,
especially if it is long and has a lot of indices. In order to make it
easier, a {\matfn Mathematica}-\TeX\ interface was written in C. It
converts the
output into a \TeX\ file using conversion rules given by the user in
a separate file. The {\funfn dvi} file produced from the \TeX\ file is
then
shown on a screen using {\funfn xdvi}, or, simply, saved.
It is necessary to work under X-Windows and have {\funfn xdvi} program
in order to show the file on a screen; otherwise, the {\funfn dvi} file
can be simply processed and printed out.

The main body of this article is organized as follows: Section 2 gives a
simple overview of SUSY and shows the kind of calculations
that can be done on a computer. Section 3 gives the algorithms for
specific manipulations and Section 4 explains the way the algorithms are
implemented in {\matfn Mathematica} software and gives information
about the package for a potential user. Examples are shown in the
Appendices.


\section{A Brief Introduction to Supersymmetry Calculations}

The algorithm and the package were developed for use in a
SUSY calculations at the classical level. The
application on a model other than $4-D,\ N=1$ requires some
definitions to be changed and the addition of model-dependent
manipulations. Nevertheless, it is general enough to include other
calculations involving similar objects and concepts. In
this section, after a brief overview of SUSY in general and some
additional properties specific to $4-D,\ N=1$ it is
explained what the calculations are, in general, that can be done
using Dill.

The usual approach to the SUSY is the superspace formulation (see, for
example \cite{Wess}).
Superfields are fields (chiral, vector, ... ) on a superspace which
has both ordinary (commuting) space-time coordinates, denoted by $x^i$,
and Grassman (anticommuting) coordinates, denoted by
$\theta^{A,\alpha}$ and the complex conjugate ${\bar \theta^{A, \dot
\alpha}}$. Here, i is a space-time index, runs from 0 to $D-1$ (where
$D$ is the space-time dimension), A from 1 to N and each of the spinor
indices: $\alpha$ and $\dot
\alpha$, from 1 to the half of the dimension of the spinor
representation of the corresponding super-Poincare algebra. (For the
dimensions in different models see for example \cite{Cornwell}.) For
$N=1$
SUSY in 4 dimensions, both $\alpha$ and $\dot \alpha$ can be 1 or 2.
By convention, a (un)dotted spinor index corresponds to a Grassman
coordinate with(out) a bar.

A superfield can be expanded in a Taylor series with respect to the
Grassman coordinates. Since the coordinates are anticommuting,
\begin{equation}
\theta^\alpha\ \theta^\beta = -\theta^\beta\ \theta^\alpha,
\end{equation}
it follows that
\begin{equation}
\theta^{\alpha}\ \theta^{\alpha} = 0
\qquad\hbox{ (no summation over}\ \alpha),
\end{equation}
so that Taylor expansion in Grassman coordinates has a finite number of
terms. The coefficients in the expansion depend on space-time only, so
they are ordinary fields. There are different superfields; for
example, the expansion of the chiral
superfield $\Phi(x, \theta)$ looks like
\begin{eqnarray}
\Phi(x, \theta) = A+ 2^{1/2}\ {\theta^{\alpha_{}}}\
{\psi_{\alpha_{}}}+ F\ {\theta^{\alpha_{}}}\ {\theta_{\alpha_{}}}+ i\
\bigl({\partial_{m_{}}}A\bigr)\ {\theta^{\alpha_{}}}\
{\sigma^{m_{}}{}_{\alpha_{}}{}_{\dot \alpha_{}}}\ {\bar \theta}{^{\dot
\alpha_{}}}+ \nonumber \\ +({\partial_{m_{}}}{\partial_{n_{}}}A)\
{\eta^{m_{}}{}^{n_{}}}\ {\theta^{\alpha_{}}}\ {\theta_{\alpha_{}}}\
{\bar \theta}{_{\dot \alpha_{}}}\ {\bar \theta}{^{\dot \alpha_{}}}/4-
i\ {\theta^{\alpha_{}}}\ {\theta_{\alpha_{}}}\
({\partial_{m_{}}}{\psi^{\beta_{}}})\ {\bar \theta}{^{\dot \beta_{}}}\
{\sigma^{m_{}}{}_{\beta_{}}{}_{\dot \beta_{}}}/2^{1/2},
\end{eqnarray}
\noindent
while the vector superfield is
\begin{eqnarray}
V(x,\theta) =
C+
i\ {\theta^{\alpha_{}}}\ {\chi_{\alpha_{}}}-
i\ {\bar \theta}{_{\dot \alpha_{}}}\ {\chi^{\dot \alpha_{}}}+
i\ {\theta^{\alpha_{}}}\ {\theta_{\alpha_{}}}\ {\bar \theta}{_{\dot
\alpha_{}}}\ ({\bar \lambda}{^{\dot \alpha_{}}}+
i\ {\bar \theta}{^{\dot \alpha_{}}}\
{\sigma^{m_{}}{}_{\beta_{}}{}_{\dot \alpha_{}}}\
({\partial_{m_{}}}{\chi^{\beta_{}}})/2)+  \nonumber \\
+i\ {\theta^{\alpha_{}}}\ {\theta_{\alpha_{}}}\ (M+
i\ N)/2-
i\ {\bar \theta}{_{\dot \alpha_{}}}\ {\bar \theta}{^{\dot \alpha_{}}}\ (M-
i\ N)/2-
{\theta^{\alpha_{}}}\ {\sigma^{m_{}}{}_{\alpha_{}}{}_{\dot
\alpha_{}}}\ {\bar \theta}{^{\dot \alpha_{}}}\ {v_{m_{}}}- \\
-i\ {\bar \theta}{_{\dot \alpha_{}}}\ {\bar \theta}{^{\dot \alpha_{}}}\
{\theta^{\alpha_{}}}\ ({\lambda^{\alpha_{}}}+
i\ {\sigma^{m_{}}{}_{\alpha_{}}{}_{\dot \beta_{}}}\
({\partial_{m_{}}}{\bar \chi}{^{\dot \beta_{}}})/2)+
{\theta^{\alpha_{}}}\ {\theta_{\alpha_{}}}\ {\bar \theta}{_{\dot
\alpha_{}}}\ {\bar \theta}{^{\dot \alpha_{}}}\ (D+
{\partial_{m_{}}}\ {\partial_{k_{}}}\ C\
{\eta^{m_{}}{}^{k_{}}}/2)/2, \nonumber
\end{eqnarray}
\noindent
where summation over the same index, appearing once as a lower and at
another place as an upper index, is assumed, as everywhere else
in this paper unless otherwise explicitly stated. $A,\ F,\ C,\ M\>
\hbox{and}\ N$
are commuting, while
$\psi,\ {\bar \psi},\ \lambda,\ {\bar \lambda},\ \chi\ \hbox{ and }
{\bar \chi}$
are anticommuting fields. By convention the $\sigma$ matrices are
given by:
\begin{equation}
\sigma^0 = \pmatrix{-1 & 0 \cr 0 & -1 \cr },\qquad
\sigma^1 = \pmatrix{0 & 1 \cr 1 & 0 \cr },\qquad
\sigma^2 = \pmatrix{0 & -i \cr i & 0 \cr },\qquad
\sigma^3 = \pmatrix{1 & 0 \cr 0 & -1 \cr}.
\end{equation}
The metric is: $\eta =\{-,+,+,+\}$ and the spinor (Grassman) indices
are raised and lowered as
\begin{equation}
\psi^\alpha = \epsilon^{\alpha \beta}\psi_\beta,\qquad
\psi_\alpha = \epsilon_{\alpha \beta}\psi^\beta,
\end{equation}
where $ \epsilon_{2\; 1} = \epsilon^{1\; 2} = 1,
\epsilon^{2\; 1} = \epsilon_{1\; 2} = -1,
\epsilon_{1\; 1} = \epsilon^{2\; 2} = 0. $
The convention used here is the one used in \cite{Wess}; different
authors may use other conventions.

Another important concept is that of a Grassman derivative which is also
anticommutative, and defined as follows:
\begin{equation}
{\partial \over \partial \theta^\alpha}\;\theta^\beta =
\delta_\alpha^\beta,\qquad
{\partial \over \partial \theta^\alpha}\;\theta^\beta\;\theta^\gamma =
({\partial \over \partial \theta^\alpha}\;\theta^\beta)\; \theta^\gamma -
\theta^\beta\;({\partial \over \partial \theta^\alpha}\;\theta^\gamma).
\end{equation}
Finally, the covariant derivatives are:
\begin{displaymath}
D_\alpha =  {\partial \over \partial \theta^\alpha} +
i\; \sigma^n_{\alpha\, {\dot \alpha}}\; {\bar \theta}^{\dot \alpha}\;
{\partial \over \partial x^n}
\end{displaymath}
\begin{equation}
{\bar D}_{\dot \alpha} =
- {\partial \over \partial {\bar \theta}^{\dot \alpha}} -
i\; \theta^\alpha\;
\sigma^n_{\alpha\, {\dot \alpha}}\; {\partial \over \partial x^n}
\end{equation}

It is necessary to see what kind of calculations are
needed in a SUSY theory and what the basic steps
needed in such a calculation are.
Typically, the calculation involve an application of the covariant
derivative on a superfield, or multiplication of two or more
superfields and sorting out the result. For the software
that can
reproduce such a calculation to be developed, the steps have to be
clearly
identified and put in an algorithmic form. Except when noted
otherwise, the steps that follow in this section are used not only in
$4-D,\ N=1$, but in
other SUSY theories also. Furthermore, the similar, if not the same,
steps are used in any other calculation where both commuting and
anticommuting tensor fields are present, so that the algorithm and the
package, with possibly minor modifications, can be applied to many
different calculations.

The important point to keep in mind is that if a package of
this kind is to be
written in a language of a symbolic programming system, the steps
that the system does and does not do by itself (without additional
programs) have to be identified. The latter are the steps that have to
be programmed in the package. For example, what the system does by
itself is:
\begin{equation}
a\ \theta^\alpha + b\ \theta^\alpha = (a + b)\ \theta^\alpha
\end{equation}

On the other hand, here are few very trivial steps that the system
does not do, so a procedure has to be written for each of them:
\begin{equation}
A\ \psi^\alpha = \psi^\alpha\ A, \qquad
\psi^\alpha\ \theta^\beta = -\theta^\beta\ \psi^\alpha,\qquad
\theta^\alpha\ \theta^\alpha = 0,\qquad
\end{equation}
\begin{equation}
\theta^\alpha\ \theta^\beta\ \theta^\gamma = 0\qquad
 ( \hbox{for}\ 4-D,\ N=1\ \hbox{ only}),
\end{equation}
where $A$ is a commuting field and $\psi$ and $\theta$ are anticommuting
fields. The multiplication has to be defined as noncommutative.
Otherwise, the system would interchange anticommuting fields in a
product without putting the necessary minus sign. The widely used
summation convention introduces a trivial step from a human point of
view:
\begin{equation}
\theta^\alpha\ \psi_\alpha = \theta^\beta\ \psi_\beta.
\end{equation}
In other words, it is not important what the actual symbol for a
summation (dummy) index is. Nevertheless, the step requires a complicated
procedure if it is to be implemented in a program.

Another set of rules comes from possible symmetries with respect to
interchange of two or more indices. If $R_{\alpha, \beta}$ is
symmetric and $B_{\alpha, \beta}$ is antisymmetric with respect to
interchange of the indices, apart from trivial identities:
\begin{equation}
R_{\alpha \beta}=R_{\beta \alpha},\qquad
B_{\alpha \beta}=-B_{\beta \alpha},
\end{equation}
the following rules are also used:
\begin{equation}
B_{\alpha \beta}\ R^{\alpha \beta} = 0,\qquad
R_{\alpha \beta}\ \theta^\alpha\ \theta^\beta = 0.
\end{equation}

The properties, or the explicit values of components of constant
tensors: $\sigma$ matrices, $\eta$'s and
$\epsilon$'s, have also to be taken into account, but a discussion
about it is deferred to the following sections.

\section{The Algorithm}

There are two major parts, or two major problems, which need to be
solved in developing an
algorithm and the software for SUSY calculations based on a
symbolic programming system. These parts and the algorithms are
explained in this section, while the details of the implementation
are given in the next one. The various
objects (superfields, component fields, constants) have to be defined
as well as the actions of operators (covariant, ordinary and Grassman
derivatives) on the (super)fields, depending both on the properties of
an operator and the index structure of the fields.
The other problem is how to take care of the symmetries and how to
identify the same terms in an expression;
by same we mean the terms that could differ at most by a
multiplicative constant.

The first part is a matter of specifying various definitions and
dependencies. It does not
require any special algorithm, but it is model dependent, so a
different set of definitions has to be written for each model. The
only point in this part that requires an explanation is the
decomposition of a product of two $\sigma $ matrices into terms with
definite symmetries, as shown in Example 2 of this section. The
definitions needed for $4-D,\ N=1$ SUSY are given in this package,
while, if one wants to work in $2-D,\ N=2$ SUSY for example, a different
set has to be given. The actual implementation of this part, functions
that perform the operations and the definitions are explained in
Section 4.

On the contrary to the
first problem, the second is almost model independent, but requires an
algorithm to solve it. The rest of the section is devoted to the
algorithm, but before that, something has to be said about the form of
an expression that is manipulated; specifically about two different ways
the dummy indices could be handled.

Since \Dill\ is intended to help a physicist by filling some
steps in a calculation, it has to accept an expression given in the
usual, compact, form as input and to give a modified expression in the same
format, as output. `Usual' means the way a physicist doing such a
calculation would put it, which implies the use of the summation
convention. The problem is how to handle the dummy (summation)
indices. There seem to be two different approaches: either all the
manipulations are done on an expression in the compact form, with
the dummy indices being symbols (letters), or all the
implicit sums in the input expression are first expanded, so that all
the dummy indices, given as symbols, are written
explicitly (as numbers) then the
required manipulation is done, and at the end the appropriate terms are
grouped together using the summation convention and dummy indices
again. The first, compact method, requires more complicated
procedures for the actual manipulations, such as differentiations or
symmetry manipulations and can not use the fact that some tensors,
($\sigma$ matrices, for example), are given constants, while the second,
explicit method, needs a complicated procedure that would pull all the
terms with explicit indices together, at the end of a calculation.
To make this important point clear, let's look at a few examples.

\vskip 10 pt
{\exfn Example 1}: Let $R_{\alpha \beta}$ be a symmetric tensor, so:
\begin{equation}
R_{\alpha \beta}\ \theta^\alpha \theta^\beta = 0
\end{equation}
The explicit method would proceed as follows ($4-D, N=1$):

\noindent
- expansion:
\begin{equation}
R_{\alpha \beta}\ \theta^\alpha \theta^\beta =
R_{1\, 1}\ \theta^1\ \theta^1 + R_{1\, 2}\ \theta^1\ \theta^2 +
R_{2\, 1}\ \theta^2\ \theta^1 + R_{2\, 2}\ \theta^2\ \theta^2
\end{equation}
- anticommutation rules:
\begin{equation}
= R_{1\, 2}\ \theta^1\ \theta^2 - R_{2\, 1}\ \theta^1\ \theta^2 \qquad
\end{equation}
- symmetry of R:
\begin{equation}
= R_{1\, 2}\ \theta^1\ \theta^2 - R_{1\, 2}\ \theta^1\ \theta^2
= 0 \quad
\end{equation}

The compact method requires a much more complicated algorithm (which
will be given later) which basically finds that:
\begin{eqnarray}
R_{\alpha \beta}\ \theta^\alpha \theta^\beta & = &
R_{\beta \alpha}\ \theta^\beta \theta^\alpha  \nonumber \\
& = & -R_{\alpha \beta}\ \theta^\alpha \theta^\beta,
\end{eqnarray}
so that:
\begin{equation}
R_{\alpha, \beta}\ \theta^\alpha \theta^\beta = 0.
\end{equation}

\vskip 10 pt
{\exfn Example 2}: Simplification of the expression:
\begin{equation}
\theta^\alpha\>{\sigma^n}_{\alpha\dot
\alpha}\>{\bar\theta}^{\dot\alpha}\;
\theta^\beta\>{\sigma^k}_{\beta \dot
\beta}\>{\bar\theta}^{\dot\beta}.
\label{eq:ex2}
\end{equation}
The explicit method would begin with an expansion of the left hand
side and explicit values of $k$ and $n$; say $k = 1,\ n = 3$, so the
previous expression equals:
\begin{eqnarray}
=\theta^1\>{\sigma^3}_{1\dot1}\>{\bar\theta}^{\dot1}\;
\theta^1\>{\sigma^1}_{1 \dot1}\>{\bar\theta}^{\dot1}& +\
\theta^1\>{\sigma^3}_{1 \dot1}\>{\bar\theta}^{\dot1}\;
\theta^1\>{\sigma^1}_{1 \dot2}\>{\bar\theta}^{\dot2} \  +\
\theta^1\>{\sigma^3}_{1\dot1}\>{\bar\theta}^{\dot1}\;
\theta^2\>{\sigma^1}_{2 \dot1}\>{\bar\theta}^{\dot1} &
 \nonumber \\ +\
\theta^1\>{\sigma^3}_{1 \dot1}\>{\bar\theta}^{\dot1}\;
\theta^2\>{\sigma^1}_{2 \dot2}\>{\bar\theta}^{\dot2}& + \
\theta^1\>{\sigma^3}_{1\dot2}\>{\bar\theta}^{\dot2}\;
\theta^1\>{\sigma^1}_{1 \dot1}\>{\bar\theta}^{\dot1}   +
\theta^1\>{\sigma^3}_{1 \dot2}\>{\bar\theta}^{\dot2}\;
\theta^1\>{\sigma^1}_{1 \dot2}\>{\bar\theta}^{\dot2}\  + &
\theta^1\>{\sigma^3}_{1\dot2}\>{\bar\theta}^{\dot2}\;
\theta^2\>{\sigma^1}_{2 \dot1}\>{\bar\theta}^{\dot1} \ \nonumber \\ +\
\theta^1\>{\sigma^3}_{1 \dot2}\>{\bar\theta}^{\dot2}\;
\theta^2\>{\sigma^4}_{2 \dot2}\>{\bar\theta}^{\dot2}  & +\
\theta^2\>{\sigma^3}_{2\dot1}\>{\bar\theta}^{\dot1}\;
\theta^1\>{\sigma^1}_{1 \dot1}\>{\bar\theta}^{\dot1} \  +\
\theta^2\>{\sigma^3}_{2 \dot1}\>{\bar\theta}^{\dot1}\;
\theta^1\>{\sigma^1}_{1 \dot2}\>{\bar\theta}^{\dot2} \  & \\ +\
\theta^2\>{\sigma^3}_{2\dot1}\>{\bar\theta}^{\dot1}\;
\theta^2\>{\sigma^1}_{2 \dot1}\>{\bar\theta}^{\dot1}  & +\
\theta^2\>{\sigma^3}_{2 \dot1}\>{\bar\theta}^{\dot1}\;
\theta^2\>{\sigma^5}_{2 \dot2}\>{\bar\theta}^{\dot2} \   +\
\theta^2\>{\sigma^3}_{2 \dot2}\>{\bar\theta}^{\dot2}\;
\theta^1\>{\sigma^1}_{1 \dot1}\>{\bar\theta}^{\dot1} \ & \nonumber \\
+\
\theta^2\>{\sigma^3}_{2 \dot2}\>{\bar\theta}^{\dot2}\;
\theta^1\>{\sigma^1}_{1 \dot2}\>{\bar\theta}^{\dot2}  & +\
\theta^2\>{\sigma^3}_{2 \dot2}\>{\bar\theta}^{\dot2}\;
\theta^2\>{\sigma^1}_{2 \dot1}\>{\bar\theta}^{\dot1} \   +\
\theta^2\>{\sigma^3}_{2 \dot2}\>{\bar\theta}^{\dot2}\;
\theta^2\>{\sigma^6}_{2 \dot2}\>{\bar\theta}^{\dot2}.  &
 \nonumber
\end{eqnarray}
Using the definition of $\sigma$ matrices:
\begin{equation}
=\ \theta^1\>{\bar\theta}^{\dot1}\;
\theta^1\>{\bar\theta}^{\dot2} \ +\
\theta^1\>{\bar\theta}^{\dot1}\;
\theta^2\>{\bar\theta}^{\dot1} \ +\
(-\theta^2\>{\bar\theta}^{\dot2})\;
\theta^1\>{\bar\theta}^{\dot2} \ +\
(-\theta^2\>{\bar\theta}^{\dot2})\;
\theta^2\>{\bar\theta}^{\dot1}
\end{equation}
anticommutation rules:
\begin{equation}
=\ \theta^1\>\theta^1\>{\bar\theta}^{\dot1}\>
{\bar\theta}^{\dot2} \ +\
\theta^1\>{\bar\theta}^{\dot1}\>{\bar\theta}^{\dot1}\>
\theta^2 \ -\
\theta^1\>\theta^2\>{\bar\theta}^{\dot2}\;
{\bar\theta}^{\dot2} \ -\
{\bar\theta}^{\dot1}\>\theta^2\>
\theta^2\>{\bar\theta}^{\dot2}\ =\ 0
\end{equation}
The whole procedure has to be repeated for all other values of $n$ and $k$.

Since the dummy indices are never written explicitly, the compact
method can not use the explicit form of the $\sigma$ matrices.
Instead, it looks at the product of two $\sigma$ matrices and
decomposes it into a sum of terms with definite symmetries:
\begin{equation}
\sigma^n{}_{\alpha {\dot \alpha}}\ \sigma^k{}_{\beta{}{\dot
\beta}}=
\sigma_+{}_+{}_+{}^{n,k}_{\alpha \beta {\dot \alpha} {\dot \beta}} +
\sigma_+{}_-{}_-{}^{n,k}_{\alpha \beta {\dot \alpha} {\dot \beta}} +
\sigma_-{}_+{}_-{}^{n,k}_{\alpha \beta {\dot \alpha} {\dot \beta}} +
\sigma_-{}_-{}_+{}^{n,k}_{\alpha \beta {\dot \alpha} {\dot \beta}},
\label{eq:Fierz}
\end{equation}
where the first three subscripts denote symmetry (+) or antisymmetry
($-$) with respect to the interchange of the space-time ($n,\ k$),
Grassman
($\alpha,\ \beta$) and dotted Grassman (${\dot \alpha},\ {\dot
\beta}$) indices, respectively. The other combinations do not appear
because of the properties of the $\sigma$ matrices. The following
identity:
\begin{equation}
\sigma_+{}_-{}_-{}^{n,k}_{\alpha \beta {\dot \alpha} {\dot \beta}}\ =\
{1 \over 2}\eta^{n\,k}\>
\epsilon_{\alpha\beta}\>\epsilon_{\dot\alpha \dot\beta},
\end{equation}
which comes from the definition of
$\sigma_+{}_-{}_-{}^{n,k}_{\alpha \beta {\dot \alpha} {\dot \beta}}$
will be used in a rule later on.

Actually, this is
an example of Fierz identities for $4-D,\ N=1$. They are different for
other models, but the logic is the same. Therefore, using
Eq.[\ref{eq:Fierz}], Eq.[\ref{eq:ex2}] becomes:
\begin{equation}
=\sigma_+{}_+{}_+{}^{n,k}_{\alpha \beta {\dot \alpha} {\dot \beta}}\;
\theta^\alpha\>{\bar\theta}^{\dot\alpha}\>
\theta^\beta\>{\bar\theta}^{\dot\beta}\ +\
\sigma_+{}_-{}_-{}^{n,k}_{\alpha \beta {\dot \alpha} {\dot \beta}}\;
\theta^\alpha\>{\bar\theta}^{\dot\alpha}\>
\theta^\beta\>{\bar\theta}^{\dot\beta}\ +\
\sigma_-{}_+{}_-{}^{n,k}_{\alpha \beta {\dot \alpha} {\dot \beta}}\;
\theta^\alpha\>{\bar\theta}^{\dot\alpha}\>
\theta^\beta\>{\bar\theta}^{\dot\beta}\ +\
\sigma_-{}_-{}_+{}^{n,k}_{\alpha \beta {\dot \alpha} {\dot \beta}}\;
\theta^\alpha\>{\bar\theta}^{\dot\alpha}\>
\theta^\beta\>{\bar\theta}^{\dot\beta}
\end{equation}
The symmetry manipulations (explained later on) give:
\begin{equation}
=\ \sigma_+{}_-{}_-{}^{n,k}_{\alpha \beta {\dot \alpha} {\dot \beta}}\;
\theta^\alpha\>{\bar\theta}^{\dot\alpha}\>
\theta^\beta\>{\bar\theta}^{\dot\beta}.
\end{equation}
Using the previously given rule:
\begin{equation}
=\ -\>{1 \over 2}\>\eta^{n\,k}\>
\epsilon_{\alpha\beta}\>\epsilon_{\dot\alpha \dot\beta}\>
\theta^\alpha\>{\bar\theta}^{\dot\alpha}\>
\theta^\beta\>{\bar\theta}^{\dot\beta}.
\end{equation}

It is clear, from the previous examples, that the explicit method is
more model dependent than the compact one because the
number of terms obtained in the expansion depends on the actual
dimensions of space-time and N, while the properties of the $\sigma$
matrices, which are model dependent, needed to be defined for both
the methods. As was already pointed out, the procedures for the compact
method are more complicated, but the problem in the explicit method is
how to group terms together once a calculation is done. The method
chosen here is the compact one; not only for its apparent beauty --
expressions are manipulated in a more human-like way, without
expanding the implicit sums, but also because it seems to be easier to
generalize for application to other models. In the rest of the section
the compact method is presented.

The expression should be in the form of a sum of terms, each
term being a product of different factors (fields, constants, ...).
Then, the idea is to transform each term so that the factors and
the indices are in a
previously specified, canonical, form. A term in the canonical form is
mathematically the same as the original term; it may only look
different because, in general,  the factors appear in a different order
and the dummy indices are different.  A canonical form should be
uniquely specified and any two terms that are the same, but have a
different form, have to have the same canonical form. Therefore, once
all the terms in an expression are brought to the canonical form,
the terms that are the same (apart from a multiplicative constant)
also {\it look} the same, so that the underlying symbolic system can
group or cancel them.

Let us define a few terms that will help understand the definition of
the canonical form. All symbolic indices appearing
in a given term are classified into two categories: free and dummy
indices. Free indices are those that appear only once,
while dummy ones appear twice, so that the
summation convention applies.
Also, all objects that depend on coordinates or have one or
more indices are called fields. Commuting numerical factors and constants
without indices are called constants.

First, the canonical form is defined and then the transformations
needed to bring a term into the order are presented. There are three
aspects to be considered for the canonical form: order of the factors,
order of dummy indices and symmetry considerations.

\vskip 20 pt
{\stfn The requirements for the canonical form:}

1) Factors are ordered so that constants come first followed by fields
sorted according to a user defined list of fields.

2) Symbols for the dummy indices are taken from the beginning
of the list of available indices (list of indices without free
ones), as many as needed.

3) If a word is constructed by taking all the indices from a term
in the order of their appearance, the word that comes from the term in
the canonical form appears before all others in dictionary order,
where the list of all indices corresponds to an alphabet.

4) A factor that has a symmetry with respect to an index permutation
has the indices ordered in the best way, according to the criterion
given in the previous requirement.

5) If a term equals zero due to an antisymmetry with respect to an index
permutation, the canonical form is simply zero.

\vskip 10 pt
It is obvious that the canonical form, defined by the requirements
given above, exists and is unique; the first two requirements
can always be satisfied, possibly in more than one way, and the third
and the fourth can be also always satisfied; furthermore,
they specify the canonical order uniquely due to the
properties of the dictionary order.

In order to clarify the definition, examples of canonical forms,
together with mathematically same terms not in the canonical
form and the number of the canonical form requirement that is {\it
not} satisfied are given in Table \ref{tab:canon}. $B_{\alpha\beta}$\
is antisymmetric,
$R_{\alpha\beta}$ is symmetric with respect to the interchange of
indices, and the user defined lists needed are:

list of fields: $\{\theta, \psi, B, R\}\qquad\qquad$
list of indices: $\{\alpha, \beta, \gamma, \delta \}$.

\noindent
Also, the upper indices come before the lower ones.

\begin{table}
\begin{tabular}{|c|c|c|} \hline
 canonical form & non-canonical form & unsatisfied
requirement \\
\hline
 $a\ \theta^\alpha\ $ & $\theta^\alpha\  a$ & 1 \\
 $\theta^\gamma\ \psi^\beta\ $ & $ -\psi^\beta\ \theta^\gamma\ $ & 1 \\
 $\theta^\alpha\  \theta^\beta\  \psi_\alpha\ $ & $\theta^\gamma\
\theta^\beta\  \psi_\gamma\ $ & 2 \\
 $-\theta^\beta\  \theta^\gamma\ $ & $\theta^\gamma\  \theta^\beta\ $
& 3 \\
 $-\theta^\alpha\  \theta_\alpha\  \theta^\beta\  \psi_\beta\ $ &
$\theta^\beta\  \theta^\alpha\  \theta_\alpha\  \psi_\beta\ $ & 3 \\
 $B_{\beta\gamma}\ $ & $ -B_{\gamma\beta}\ $ & 4 \\
 $-\theta^\alpha\  \theta^\beta\  B_{\alpha \beta}$ & $\theta^\alpha\
\theta^\beta\  B_{\beta \alpha}$ & 4 \\
 $ 0 $ & $ \theta^\alpha\ \theta^\beta\ R_{\alpha \beta}$ & 5 \\ \hline
\end{tabular}
\caption{Canonical ordering}
\label{tab:canon}
\end{table}

In all the examples given in Table \ref{tab:canon} only Grassman
indices appear. If
more than one type of index is needed, as is usually the case,
the canonical form definition essentially stays the same; all the
orderings are done for all index types. A list of indices has to be
defined for each type, so that the dummy indices reassignment is done
separately for each index type. The convention used here is
that (dotted) Greek letters denote (dotted) Grassman
indices and small Latin letters are used for space-time coordinates.
Although it would be mathematically correct, it would be odd to use
for an index one kind of letter, when another kind is expected.

The main steps in the procedure (algorithm) that transforms a term
into the canonical form are first listed and the explanations
given immediately after.

\vskip 20 pt
{\stfn Canonical ordering algorithm}:

1) Factors in a term are ordered so that constants come first,
followed by fields ordered according to the list of fields.

2) A set of terms is generated by making all the possible permutations
of fields within each group of the same fields.

3) Another set of terms is generated from the previous one by using all
existing symmetries with respect to index interchange of a field.

4) The dummy indices are
reassigned from a list of available indices, according to the order in
the list, for every term in the set.

5) A term from the set that has indices ordered in the best way is
chosen for the canonical form, unless it is zero due to the
symmetries, in which case zero is returned.

\vskip 10 pt
In the first step, factors are ordered so that
all commutative constants come at the beginning, followed by fields
and other anticommuting objects in a user specified order (a user
creates a list of fields). Therefore, following constants, groups of
one or more of the same fields are created. When this
is done, the orders of the anticommuting factors of the initial and
final factors are
compared, so that the parity of a permutation needed to bring one into
another is determined and if it is odd, an additional minus sign is
introduced in the final term. In this way the minus signs coming from the
anticommutation relations are taken into account properly.

There are two types of symmetries that are of concern here: symmetry
with respect to a permutation of indices of a single field (
$B_{\alpha\beta} = -B_{\beta\alpha}$, or
$T_{\alpha\beta\gamma} = T_{\beta\gamma\alpha} =
T_{\gamma\alpha\beta}$)
and with respect to interchange of the same fields
($\theta^\alpha\ \theta^\beta = -\theta^\beta\ \theta^\alpha$). They
both amount to an index permutation which makes them
important in finding a canonical form. The term
obtained from the previous step is used to generate the
terms with all possible permutations of indices that are equal to the
term, modulo minus sign. The way it is done is to take each group of the
same fields separately, make all possible field permutations within
the groups and then combine all the permutations, followed by the usual
addition of a minus sign if necessary. Now, the same thing
is done with fields that have a definite symmetry with respect to an
index permutation (the fields and the corresponding symmetries are
specified by the user). The overall sign change depends on the number of
antisymmetries used to generate the particular term. What is obtained
in this way is a set of terms  with all possible index combinations,
each equal to the original term.

In the next step of the algorithm, each term from a set is treated
separately, so the result will be another set of terms, again each
term from the set being equal to the original one.
The free indices in a term have to stay as they are; on the other
hand, a dummy index can be given
a different symbol, provided both of the same indices are given the same,
new, symbol. A list of available indices is obtained by removing the
free indices of the actual term, from the list of
all indices. The
dummy indices are reassigned in the order they appear in
the term: no matter what the actual
symbol is, they are assigned a new one, in the order of the available
list of indices.

A set of terms obtained in general contains terms that look identical.
In order to make further manipulations faster, only distinct terms are
retained in the set. If there is more than one kind of index, the
reassignment is done with each kind separately, one after another.

What is achieved by the above procedure is that starting from an initial
term, a set of terms are generated equal to it. In every term all
the fields are in the same order.
The terms differ in positions of dummy indices; the
symbols used for the dummy indices are the same, but they are assigned
in all the possible ways following all the symmetries with respect to
permutations of indices and terms.

The final step is to find a term in the canonical form in the set.
The first two canonical form requirements are satisfied by
construction. As for the next two requirements, a word of indices in
the order of their appearance is constructed for every term, all of
them are sorted as in a dictionary following the list of all indices
and the first one corresponds to a term in the canonical form. If the
chosen word appears more than once, the terms corresponding to them can
either be identical, in which case it does not matter, or they can
differ by an overall minus sign, meaning that the terms equals minus
the same term, so it is zero.
In this case, in the last step of the algorithm, the term is set to
zero. Since all possible symmetries of a term
are taken into account for generating the set, if the term
is zero mathematically, the same terms with different signs will
necessarily appear in the set and will be set to zero.
Two examples of transforming expressions to the canonical form,
following the steps from the canonical ordering algorithm are given in
Appendices A and B.

It has thus been shown that not only does the canonical form of a term
(as given by the requirements) exist and that it is unique, but also
that the procedure given above necessarily leads to it.

\section{Implementation}

The previous discussion was mostly concerned with the algorithm and
therefore independent of the actual symbolic programming system used,
although the common characteristics of the existing systems
were taken into account. In that way, the algorithm presented above
can be implemented in anyone of them. This section is in turn devoted
to the actual implementation of the algorithm in {\matfn
Mathematica}. The functions
corresponding to specific manipulations have been written in the
{\matfn Mathematica} programming language and the use and the
function of the several most important of them is given. The actual
form of the functions was strongly influenced by the existing
{\matfn Mathematica} commands. If the software is to be written in another
similar language the details would have to be different; nevertheless
the basic structure would stay the same.

Following the ideas of object oriented programming, the
whole package is written in a highly modular form. Every function
performs a single well-defined step, so that it is easy to understand
the way a function works and it also gives \Dill\  greater
flexibility. All the functions
in the package can be roughly divided into the main and other
functions.  A user needs to be familiar with the main functions only,
which make up a small fraction of
all the functions. The main functions fall in three categories: output
commands, format changes and the functions that perform an actual
transformation on an expression. The transformations are quite
similar to those a physicist would do in an usual calculation. The
main functions are listed below. The other functions
are typically either general, therefore used
everywhere, or closely related and providing a necessary step for a
specific main function, but are not used by themselves. Unless a
user wants to change a function, there is no need to get familiar
with them, so they will not be of concern in this paper.

\vskip 10 pt {\stfn Format changing functions.}
Before a list and a short description of the functions is given, it is
necessary to give an explanation about the forms in which the
expressions actually appear during the calculations.
Instead of working with an expression in its usual form -- fields and
constants that are added and multiplied amongst themselves --
it was very convenient to establish a few different formats for an
expression. A function that performs a certain transformation requires
an input expression in a specific form. First of all, given an
expression, all multiplications (both commutative and anticommutative)
and derivatives (both ordinary and Grassman) have to be
distributed over additions, so that the expression is a sum of
a product of factors, where a factor can be a constant, a field, or
any derivative of any order of a field. This format is called
the ordinary expression format. The function {\funfn initdis}
acts on any
expression and puts it in the {\em ordinary expression} form.

If a term of an {\em ordinary expression} (a product of constants, fields
and derivatives) is written as a list (set) with the
factors as elements, the term is in the {\em list} format. The whole
expression (sum of terms) is in the {\em lists} format if it is a list of
terms, each term being in the {\em list} format. That is, an expression in
the {\em lists} form is a list of lists of constants, fields and
derivatives. This is achieved by the {\funfn oex2lists} function.
Further, indices of most of the fields -- those listed in
{\funfn raiselist} -- have to be up. The function used for this is {\funfn
raise}.

The last two formats; {\em full} for a single term and {\em fulls} for
the whole
expression are again similar: a list of terms in the {\em full} format
makes an expression in the {\em fulls} format. A term in the {\em
full} form is a
list of two lists: the first one has constants for elements. The
second list contains fields and derivatives grouped in lists of the
same fields, as shown in Example~3. The fields are ordered according to
{\funfn funlist}. The function {\funfn lists2fulls} transforms an
expression in the {\em lists} format to {\em fulls}. It is strongly
recommended
that there is no Grassman derivatives nor Kronecker delta functions in
the expression.

\vskip 10 pt
{\exfn Example 3}:

\noindent
Initial expression:
\begin{equation}
\theta^\alpha\>\theta_\alpha
\>{\bar\theta}^{\dot\alpha}\>{\bar\theta}_{\dot\alpha}\ +\
({\partial \over \partial\theta^\alpha}\,\theta^\beta\,\theta^\gamma)
\>(\psi^\alpha\,\epsilon_{\beta\gamma}
)\ -\ A\,{\partial \over \partial x^n}\,{\partial \over \partial x^k}\,
\eta^{k n}
\end{equation}
{\em ordinary expression}:
\begin{equation}
\theta^\alpha\>\theta_\alpha
\>{\bar\theta}^{\dot\alpha}\>{\bar\theta}_{\dot\alpha}\ +\
 ({\partial \over \partial\theta^\alpha}\,\theta^\beta)\>\theta^\gamma\>
\psi^\alpha\,\epsilon_{\beta\gamma}
\ -\ \theta^\beta\>({\partial \over\partial\theta^\alpha}\,\theta^\gamma)
\>\psi^\alpha\,\epsilon_{\beta\gamma}
\ +\ A\,({\partial \over \partial x^n}\,{\partial \over \partial x^k}\,A)
\>\eta^{k n}
\end{equation}
{\em lists} form:
\begin{displaymath}
\{
\{\theta^\alpha,\ \theta^\beta,\ \epsilon_{\alpha\beta},
\ {\bar\theta}^{\dot\alpha},\ {\bar\theta}_{\dot\alpha}
,\ \epsilon_{{\dot\alpha}{\dot\beta}}\},\
\ \{ {\partial \over\partial\theta^\alpha}\ \theta^\beta,
\ \theta^\gamma,\
\psi^\alpha,\ \epsilon_{\beta\gamma}\},\
\end{displaymath}
\begin{equation}
\{-1,\ \theta^\beta,\ {\partial \over\partial\theta^\alpha}\,\theta^\gamma,
\ \psi^\alpha,\ \epsilon_{\beta\gamma}\},\
\{A,\ {\partial \over \partial x^n}\,{\partial \over \partial x^k}\,A,
\ \eta^{k n}\} \
\}
\end{equation}
{\em fulls} form:
\begin{displaymath}
\{
 \{
\ \{1\},
\ \{
\{
\epsilon_{\alpha \beta}
\}
,\{
{\bar\epsilon}_{{\dot\alpha} {\dot\beta}}
,\},
\{
\theta^\alpha,\>\theta^\beta
\},
\{
{\bar\theta}^{\dot\alpha},
{\bar\theta}^{\dot\beta}
\}
\}
\ \},\
\{\
\{
1
\},\
\{
\{
{\partial \over \partial\theta^\alpha}\,\theta^\beta
\},
\ \{
\theta^\gamma
\},
\ \{
\epsilon_{\beta\gamma}
\},
\{
\psi^\alpha
\}
\}
\ \},
\end{displaymath}
\begin{equation}
\{
\ \{
-1
\},
\ \{
\{
{\partial \over\partial\theta^\alpha}\,\theta^\gamma
\},
\{
\theta^\beta
\},
\{
\epsilon_{\beta\gamma}
\},
\{
\psi^\alpha
\}
\}
\ \},\
\{
\{
1
\},
\ \{
\{
{\partial \over \partial x^n}\,{\partial \over \partial x^k}\,A
\},
\{
A
\},
\{
\eta^{k n}
\}
\}
\ \}\
\}
\end{equation}

In all the following function descriptions, the argument {\funfn ind}
stands for an index type, {\funfn indtypes} for a list of index types,
{\funfn funlist} for a list of fields and
{\funfn oddlist} for a list of fields that have odd Grassman parity.

The format changing functions are the following:

- {\funfn initdis[ ex, oddlist ]} distributes derivatives,
ordinary and noncommutative multiplications over additions in the
expression {\funfn ex} and returns the expression in the ordinary
expression format.

 - {\funfn raise[ term, indtypes, raiselis ]} takes a term in the {\em
list}
form and raises all the indices of the types {\funfn indty} in the
factors
appearing in {\funfn raiselis}.

 - {\funfn oex2lists[ oex ]} transforms an ordinary form
expression {\funfn oex} to the {\em list} form.

 - {\funfn lists2oex[ ex ]} transforms an expression in the {\em lists}
form to the {\em ordinary expression} form.

 - {\funfn list2oex[ term ]} transforms a term in the {\em list} form
to an expression in the ordinary form.

 - {\funfn lists2fulls[ ex, funlist, oddlist ]} transforms an
expression in the {\em lists} form to the {\em fulls} form.

 - {\funfn list2full[ term, funlist, oddlist ]} transforms a term
in the {\em list} form to the {\em full} form.

 - {\funfn fulls2lists[ ex ]} transforms an expression in the
{\em fulls} form to the {\em lists} form.

 - {\funfn full2list[ term ]} transforms a term in the {\em full} form
to the {\em list} form.

 - {\funfn oex2fulls[ oex, funlist, oddlist]} transforms an
{\em ordinary} form expression to the {\em fulls} form.

 - {\funfn fulls2oex[ ex ]} transforms an expression in the {\em fulls}
form to the {\em ordinary expression} form.

\vskip 10 pt {\stfn Output commands} are not necessary but are very
convenient. An expression manipulation or a format changing function
returns an expression in the {\matfn Mathematica} expression form (all
of the formats discussed above agree with the {\matfn Mathematica}
expression form), so that the output expression can be used as an
argument of a {\matfn Mathematica} command. Unfortunately it is not
easy for a user to read it. Especially in
the case of the interactive use, when a user decides about the next
step depending on the actual expression, it seems very important to
develop a way to display an expression in the same way it would be
written by a physicist. For example, the way \Dill\ is used,
the expression:
\begin{equation}
\theta^\alpha\>{\sigma^n}_{\alpha\dot
\alpha}\>{\bar\theta}^{\dot\alpha}\;
\theta^\beta\>{\sigma^k}_{\beta \dot
\beta}\>{\bar\theta}^{\dot\beta}
\end{equation}
would be in the {\matfn Mathematica} form:

\noindent
\centerline{
theta[$\{$alpha, u$\}$] ** sigma[$\{$n, u$\}$,\ $\{$alpha, d$\}$,
$\{$alphad, d$\}$] **
thetab[$\{$alphad, u$\}$]
\ **
}

\noindent
\centerline{
theta[$\{$beta, u$\}$] **
sigma[$\{$k, u$\}$, $\{$beta, d$\}$, $\{$betad, d$\}$] **
thetab[$\{$betad, u$\}$].
}

\vskip 10 pt
There is a way to display subscripts and superscripts in an expression
built into {\matfn Mathematica}, but it is not flexible enough to use
in this package. In order to allow the presentation of an expression
in an user -- friendly
way and to make it easy to save expression in a file, output commands
are developed. A user has to specify whether the output should be
displayed in \TeX\ format, or in {\matfn Mathematica} format, should
be sent to
the standard output or to a file, and, in the case that the standard
output (usually a screen) and \TeX\ form is chosen, should the results
of different manipulations be displayed one by one, as they are
calculated, or all together, when the whole job is done. In any case,
the output commands affect only the way a user gets a result;
{\matfn Mathematica} itself always gets the output in {\matfn
Mathematica} form, and is therefore ready for another
manipulation. Further, in a program that does
several manipulations, the output commands can be used
anywhere in the program and, at the time the program is run, the user
can specify if {\funfn all} of them are to be used, {\funfn some}, or
the {\funfn last} one only.

The output commands are:

 - {\funfn outopen[ outdes ]} defines parameters needed for
{\funfn writework} and {\funfn outclose}. The argument is a list of
four elements: the first can be {\funfn all, some}, or {\funfn last},
for all the second {\funfn ``stdout''}, or a {\funfn ``filename''},
the third
{\funfn mat}, or {\funfn tex} and the fourth {\funfn tog}, or {\funfn
one} (all outputs are put together, or appear one by one). Returns a
list of three elements.

 - {\funfn writework[ out, string, ex ]} sends the expression
{\funfn ex} to the output preceded by a word {\funfn string} enclosed
in quotes. The argument {\funfn out} is one of the three elements of
the list returned by {\funfn outopen}. {\funfn writework} with the
argument {\funfn out}
being the third element of the list should be used to print the
end-result only, writework with the second element of the list should
be called after an important intermediate result is generated
(generates a
short output) and with the first, after each step in the calculation
(used for a long output).

 - {\funfn outclose[ outdes ]} does whatever is needed to finish
the output processes started by {\funfn outopen}. The arguments are
the same as for {\funfn outopen}.

 - {\funfn writex[ ex ]} displays the expression {\funfn ex} on
the screen as it would be printed by \TeX.

 - {\funfn mrep} is a C program that takes a file with
{\matfn Mathematica} expressions (whose name is the first argument)
and transforms it into a \TeX\ file (whose name is the second
argument), by making pattern substitutions defined in the substitution
file provided by the user (whose name is the third argument; {\funfn
wb.sub} is the substitution file in the package). A
pattern/substitution pair can include wild cards. Additional
information about {\funfn mrep} and the format of a substitution file
are provided in the file {\funfn mrep.c} and in the other source files.

 - {\funfn ftxdvi} is a shell script that performs {\funfn
mrep}, \TeX\ the result and prints it on a screen using the {funfn
xdvi} command. The first argument is the name of the input file and
the second is the name of the substitution file. The input file and the
files generated in the process; a \TeX\ and a {\funfn dvi} file, have
the same names; only the extensions are changed appropriately.

 - {\funfn ftx} is a shell script similar to {\funfn ftxdvi}. The
only difference is that it does not display the output on a screen.

\vskip 10 pt {\stfn Expression manipulation functions.} These
functions perform actual calculations, so they are the core of the
package. The functions are:

 - {\funfn dman[ ex ]} applies the usual rules for Grassman and
ordinary derivatives on the expression {\funfn ex} in the lists
format. {\funfn deplist[x]} is needed to specify fields that depend on
the space-time coordinate x.

 - {\funfn metricman[ term, indtypes ]} takes a {\em list} form term
{\funfn term} and simplifies products of metrics (index raising
function) with upper and lower indices of the types listed in the list
{\funfn indtypes}.

 - {\funfn deltaman[ term, indtypes ]} takes care of the Kronecker
delta function with indices of the types {\funfn indtypes} in the {\em
list} form term {\funfn term}.

 - {\funfn toodd[ term, ind, oddlist ]} multiplies a {\em full} form
term
{\funfn term} by zero if it contains as many anticommuting factors
with
indices of the {\funfn ind} type as needed for the term to be zero.

 - {\funfn zero[ ex ]} deletes the terms from the {\em fulls} form
expression {\funfn ex} that have a zero factor.

 - {\funfn ssigma[ ex, funlist, oddlist ]} decomposes a product
of two $\sigma$ matrices into the terms with definite symmetries in
the {\em fulls} form expression {\funfn ex}.

 - {\funfn canon[ term, out, funlist, oddlist ]} puts a {\em full}
form
term into the canonical order and prints the steps if {\funfn all} was
an argument to {\funfn outopen}. The argument {\funfn out} is the
whole output of {\funfn outopen}. It has to be noted that {\funfn
canon} does not order the factors in the term (step 1 in the canonical
order algorithm) because it expects the input ({\funfn term}) in the
{\em full} form, so it is ordered already.

 - {\funfn work[ ex, outdes, funlist, oddlist ]} is a program that
applies the functions described above, in a certain order, to the input
expression {\funfn ex}. The argument {\funfn outdes} is the argument
needed for {\funfn outopen} and {\funfn outclose}. The function is
listed and its use is shown in the Appendix~D.

Since {\funfn canon} is the most important function in the package -- it
is an implementation of the algorithm given in Section 3 -- the
functions explicitly called in {\funfn canon} are listed as follows:

 - {\funfn permfac[ term, oddlist ]}: makes all possible
permutations of the same fields in a {\em full} form term {\funfn term},
assigns a sign as needed due to the anticommutation relations and
returns a list of terms with the fields permuted; effectively an
expression in the {\em fulls} form. (Step 2 in the canonical ordering
algorithm.)

 - {\funfn symall[ term, ind ]}: makes all the permutations of
indices of the type {\funfn ind}, according to the (anti)symmetries
given by {\funfn symtype[ field, ind ]} lists, in the {\em full} form
term {\funfn term} and returns a list of the transformed terms. (Step
3 in the canonical ordering algorithm.)

 - {\funfn shuffle[ term, ind[b] ]}: reassigns dummy indices of
the type {\funfn ind} in the {\em full} form term {\funfn term} in the
order of their appearance, according to the list of indices with the
free indices excluded. (Step 4 in the canonical ordering algorithm.)

 - {\funfn firstbyind[ ex, indlist ]} takes a list of {\em full} form
terms {\funfn ex}, all of them mathematically the same and returns the
list of the terms with the best alphabetical ordering of the indices,
according to the list of indices {\funfn indlist} (there may be more
than one term with the same  index order). If there is more than one
term and the terms have different signs a zero is appended to the
constant
list, effectively setting the term to zero. (Step 5 in the canonical
ordering algorithm.)

Two applications of the functions appearing in {\funfn canon} are
shown in Appendices A and~B.

\vskip 10 pt {\stfn Definitions and declarations}.
Also needed are different definitions and declarations:

 - {\funfn funlist}: a list of all fields.

 - {\funfn oddlist}: a list of all fields with odd Grassman
parity.

 - {\funfn deplist[x]}: a list of all fields that depend on
space-time.

The following items need to be defined for all index types:

 - {\funfn raiselist[indextype]} a list needed in {\funfn raise}.

 - {\funfn ind[ ]} a list of all indices of the type {\funfn ind}.
An index has a form of a list where the first element is a symbol for
the index and the second is {\funfn u}, or {\funfn d} denoting up or
down
position of the index.
Must not have any element in common with a similar list for another
index type.

 - {\funfn ind[d]}: a list of all up indices of the type {\funfn
ind}.

 - {\funfn ind[u]}: a list of all down indices of the type {\funfn
ind}.

 - {\funfn ind[b]}: a list of all index symbols, without the
position, of the type {\funfn ind}.

 - {\funfn ind[ch]}: a function that raises the indices of the type
{\funfn ind}.

 - {\funfn ind[e]}: a list of the explicit index values
(integers) of the type {\funfn ind}. (The only thing needed in the
current implementation is the length of the list, that is the
dimension of the (super)space related to the indices.)

The following lists need to be defined for all field and index type
combinations:

 - {\funfn symtype[ field, ind ]}: a list of all symmetries
of {\funfn field} with respect to a permutation of indices of the type
{\funfn ind}. Every element of the list is a list; the first element
is: $\{1\}$ , or $\{-1\}$ for the symmetry or the antisymmetry, and the
second is a list denoting the permutation.

Different objects, superfields or covariant derivatives for example,
have to be defined if needed. Care has to be taken if an object
contains dummy indices within its definition, as is usually the case,
because the dummy indices must not have the same symbol as the other
indices appearing in the expression containing the object. All the
indices appearing in the expression, the dummy indices of the object
excluded, have to be listed first, and then the dummy indices can be
assigned some other symbols. It would be possible to do it by hand, or
to assign the dummy indices from a special set of index symbols not
used elsewhere, but it would be very impractical, especially when an
expression contains more than one object of the same kind; the dummy
indices in all the objects must be different. In order to achieve this
a function {\funfn wbfields} is created. It contains definitions for
covariant derivatives, vector and scalar fields for $N = 1,\  4-D$ case:

 - {\funfn wbfields}[ oex ] substitutes objects defined in the
function by the definitions, in an {\em ordinary expression} {\funfn
oex},
in such a way that all the dummy indices are different.
For other definitions, {\funfn wbfields} should be used as a template.
The name can be changed, or other definitions can be added in the
same way the existing ones appear.

The functions can be applied one by one (interactively), or combined
in a program (a noninteractive case). In the later case, the functions
would be
applied one after another in a given order.
The function {\funfn work} is an example of the program used in the
noninteractive case.
The expressions obtained
after different steps can be easily shown together, or written in a
file. This approach is convenient if the calculation takes a long
time, but the order of required manipulations is known in advance.
Examples of a simplification of an expression using the functions
listed above, are given in Appendix~C for the interactive and in
Appendix~D nor the noninteractive cases respectively.

\vskip 10 pt {\stfn Other important remarks}.
Although, as
can be seen from the previous function descriptions, several
functions that perform a manipulation have as argument not a whole
expression, but a term only, it is easy to apply the same function on
a whole expression. An expression in both {\em lists} or {\em fulls}
form is a list of terms in the list or the {\em full} form:

expression = $\{$ term\_1, term\_2, ... $\}$

\noindent
so if {\funfn function} takes  a term for an argument:

{\funfn function}[ term, ... ],

\noindent
where ... stands for other arguments, the construction:

{\funfn function}[ \#, ... ]\& /@ expression

\noindent
gives

$\{$ {\funfn function}[ term\_1, ... ], {\funfn function}[ term\_2,
... ],
... $\}$

\noindent
that is, it applies {\funfn function} on each term and puts the
results in a list, thereby giving the output expression in the same
form as the input one. This is important if another function has to be
applied to the output expression.

Another important remark concerns the derivatives. While in
{\matfn Mathematica}

\hskip 1 in
$D[ f, x ]\qquad \hbox{stands for:}\qquad {\partial \over
\partial\,x}\,f$,

\noindent
in this package:

\hskip 1 in
${\partial \over \partial\,x^n}\,f \qquad
\hbox{ is written as:}\qquad d[ x[\{n, u\}], f ]$.

More detailed instructions are provided in the package; they
are integrated in the help utility of {\matfn Mathematica} and can be
obtained in the same way as information about any other command.


\section{Conclusion}

An algorithm for doing SUSY calculations at the classical
level consists of two major parts. The application of operators,
definitions and standard rules, such as Fierz identities and Kronecker
delta function properties, belong to the first part. The second part
relates to transformation of the terms of an expression to a previously
specified canonical form using all the symmetry properties of the
fields involved, so that when the algorithm is implemented, the
underlying symbolic programming system can identify the terms with the
same form and group them together or cancel them. The number of terms
in a typical supersymmetry calculation and the complicated structure
involving different symmetries and summations over many different
indices makes the second part much more important. An algorithm for
transforming a term into its canonical form, together with a
definition of the canonical form, was found and presented. The
algorithm is model independent and it was shown that the canonical
form was unique and that the algorithm necessarily lead to it.

The complete algorithm (both parts), for $4-D,\ N=1$ SUSY theory, was
implemented in the {\matfn Mathematica} programming language. \Dill ,
the software package obtained
this way, consists of functions that perform different manipulations of
the first part, a function that brings a term into the canonical form,
as well as other functions needed for the proper functioning of the main
functions and output commands. The package was tested on several
examples, only a few of which are presented in this article. An
initial expression
was supplied, the appropriate transformations were applied to it and
the final result was obtained. It was found that \Dill\ always
gives the correct
result. In some cases, as shown in Appendix~D, the initial expression
was not simplified to the fullest extent, because a rule particular to
$4-D,\ N=1$ model was not supplied. Nevertheless, the result obtained
is mathematically correct.
Therefore, although the package in its current form
does not perform all possible manipulations on an expression, the
manipulations that it does perform are done correctly.
\Dill\ is intended to be an interactive tool
(although it can be also used as an ordinary program) and an open set
of functions, if another manipulation is needed a function that does
it can simply be added. Different models would require some functions
and definitions to be changed, but the overall structure and the
canonical ordering function can stay the same. The software was run on
Sun SPARC-10 workstations. The CPU time needed was
typically somewhat bigger, but of the same order than the time an
average physicist would need; 10 min compared to 5 min for example.

There are several things that could be done to improve the package. It
is always possible to make it faster; the functions were written with
the main motivation of working properly. In the course of developing
the package a lot of them were simplified and made faster, but there
is still space for additional improvements. Another idea would be to
incorporate the explicit method explained in Section~3 and use it
together with currently used compact method. That would allow simpler
implementations of model dependent rules, such as the one needed for
the complete simplification in Appendix~D.

Furthermore, \Dill\ can be extended to include gauge fields and
superfields, making it suitable for calculations in super Yang-Mills
theories. Apart from additional declarations and definitions, the
extension would require rules related to commutators of the Lie
algebra used in the theory.

The package is available for free distribution. For the further
information please contact the author.

\vfill \eject

\appendix
\abvt
\noindent
{\Large\bf Appendix A: Canonical Ordering Procedure, Example 1}
\belt

In this appendix, the expression:
$\theta^\gamma\,\psi^\beta\,\epsilon_{\delta \gamma}\,\theta^\delta$
is transformed according to the canonical ordering procedure given in
Section 3. Functions explicitly appearing in {\funfn canon} (see
Section 4), that correspond to the steps in the procedure are shown in
the way they would be typed by the user.

\noindent
Initial expression:

\centerline{$
\theta^\gamma\,\psi^\beta\,\epsilon_{\delta \gamma}\,\theta^\delta
$}

\noindent
Step 1 of the canonical ordering algorithm:

\centerline{
$-\theta^\gamma\,\theta^\delta\,\epsilon_{\delta \gamma}\,\psi^\beta$}

\noindent
Step 2 of the canonical ordering algorithm:

Corresponding command:
{\funfn ex = permfac[ exps, oddlist ];
}

\vskip -15 pt
$$
\{\ -\theta^\gamma\,\theta^\delta\,\epsilon_{\delta
\gamma}\,\psi^\beta, \qquad
\theta^\delta\,\theta^\gamma\,\epsilon_{\delta \gamma}\,\psi^\beta\
\}$$

\noindent
Step 3 of the canonical ordering algorithm:

Corresponding command:

{\funfn
      ex = Flatten[ symall[ \#, lor ]\& /@ ex, 1 ];

      ex = Flatten[ symall[ \#, gras ]\& /@ ex, 1 ];

      ex = Flatten[ symall[ \#, grasd ]\& /@ ex, 1 ];
}

\vskip -15 pt
$$
\{\ -\theta^\gamma\,\theta^\delta\,\epsilon_{\delta
\gamma}\,\psi^\beta,\qquad
\theta^\delta\,\theta^\gamma\,\epsilon_{\delta \gamma}\,\psi^\beta,
\qquad \theta^\gamma\,\theta^\delta\,\epsilon_{\gamma
\delta}\,\psi^\beta, \qquad
-\theta^\delta\,\theta^\gamma\,\epsilon_{\gamma \delta}\,\psi^\beta\
\}$$

\noindent
Step 4 of the canonical ordering algorithm:

Corresponding command:

{\funfn       ex = shuffle[ \#, lor[b] ]\& /@ ex // Union;

      ex = shuffle[ \#, gras[b] ]\& /@ ex // Union;

      ex = shuffle[ \#, grasd[b] ]\& /@ ex // Union;
}

\vskip -15 pt
$$\{\ -\theta^\alpha\,\theta^\gamma\,\epsilon_{\gamma
\alpha}\,\psi^\beta,\qquad
\theta^\alpha\,\theta^\gamma\,\epsilon_{\alpha \gamma}\,\psi^\beta,
\qquad \theta^\alpha\,\theta^\gamma\,\epsilon_{\alpha
\gamma}\,\psi^\beta, \qquad
-\theta^\alpha\,\theta^\gamma\,\epsilon_{\gamma \alpha}\,\psi^\beta\
\}$$

\noindent
Step 5 of the canonical ordering algorithm:

Corresponding command:

{\funfn       exps = firstbyind[ex, Join[lor[], gras[], grasd[]] ][[1]];
}

\vskip -15 pt
$$ \theta^\alpha\,\theta^\gamma\,\epsilon_{\alpha
\gamma}\,\psi^\beta$$

\abvt
\noindent
{\Large\bf Appendix B:  Canonical Ordering Procedure, Example 2}
\belt

In this appendix, the canonical ordering procedure is applied to the
expression:

\noindent
\centerline{
${R_{\gamma_{}}{}_{\beta_{}}}\,{\theta^{\beta_{}}}\,{\theta^{\gamma_{}}}$
}

\noindent
Step 1 of the canonical ordering algorithm:

\noindent
\centerline{
${\theta^{\beta_{}}}\,{\theta^{\gamma_{}}}\,{R_{\gamma_{}}
{}_{\beta_{}}}$ }

\noindent
Step 2 of the canonical ordering algorithm:

Corresponding command:
{\funfn ex = permfac[ exps, oddlist ];
}

\vskip -15 pt
$$\{\ {\theta^{\beta_{}}}\,{\theta^{\gamma_{}}}\,{R_{\gamma_{}}
{}_{\beta_{}}}, \qquad -
{\theta^{\gamma_{}}}\,{\theta^{\beta_{}}}\,{R_{\gamma_{}}
{}_{\beta_{}}}\ \}$$

\noindent
Step 3 of the canonical ordering algorithm:

Corresponding command:

{\funfn
      ex = Flatten[ symall[ \#, lor ]\& /@ ex, 1 ];

      ex = Flatten[ symall[ \#, gras ]\& /@ ex, 1 ];

      ex = Flatten[ symall[ \#, grasd ]\& /@ ex, 1 ];
}

\vskip -15 pt
$$\{\ {\theta^{\beta_{}}}\,{\theta^{\gamma_{}}}\,{R_{\gamma_{}}
{}_{\beta_{}}},\qquad
{\theta^{\beta_{}}}\,{\theta^{\gamma_{}}}\,{R_{\beta_{}}
{}_{\gamma_{}}},\qquad -
{\theta^{\gamma_{}}}\,{\theta^{\beta_{}}}\,{R_{\gamma_{}}
{}_{\beta_{}}},\qquad -
{\theta^{\gamma_{}}}\,{\theta^{\beta_{}}}\,{R_{\beta_{}}
{}_{\gamma_{}}} \ \}$$

\noindent
Step 4 of the canonical ordering algorithm:

Corresponding command:

{\funfn       ex = shuffle[ \#, lor[b] ]\& /@ ex // Union;

      ex = shuffle[ \#, gras[b] ]\& /@ ex // Union;

      ex = shuffle[ \#, grasd[b] ]\& /@ ex // Union;
}

\vskip -15 pt
$$\{\ -{\theta^{\alpha_{}}}\,{\theta^{\beta_{}}}\,{R_{\alpha_{}}
{}_{\beta_{}}},\qquad -
{\theta^{\alpha_{}}}\,{\theta^{\beta_{}}}\,{R_{\beta_{}}
{}_{\alpha_{}}},\qquad
{\theta^{\alpha_{}}}\,{\theta^{\beta_{}}}\,{R_{\alpha_{}}
{}_{\beta_{}}},\qquad
{\theta^{\alpha_{}}}\,{\theta^{\beta_{}}}\,{R_{\beta_{}}
{}_{\alpha_{}}} \ \}$$

\noindent
Step 5 of the canonical ordering algorithm:

Corresponding command:

{\funfn       exps = firstbyind[ex, Join[lor[], gras[], grasd[]] ][[1]];
}

\vskip -15 pt
$$0\
{\theta^{\alpha_{}}}\,{\theta^{\beta_{}}}\,{R_{\alpha_{}}{}_{\beta_{}}}
\ =\ 0$$

\abvt
\noindent
{\Large\bf Appendix C: Interactive use of the package
}
\belt

In this example, the expression:

\centerline{
${D_{\gamma_{}}}\bigl({\theta^{\delta_{}}}\ {\theta_{\delta_{}}}\ F\bigr)$
}

\noindent
is evaluated using the functions from the package.
The functions needed are shown, together with the output in the \TeX\
mode. This is an example of the interactive use of the package. The
output commands that put an output expression in the \TeX\
mode are neglected.

\vskip 10 pt
\noindent
{\funfn ex = DSS[ $\{$gamma, d$\}$, theta[$\{$delta, u$\}$] **
theta[$\{$delta, d$\}$] ** F[] ]}

\noindent
{\funfn ex = wbfields[ ex ] }

\centerline{
${\partial_{\gamma_{}}}\bigl({\theta^{\delta_{}}}\
{\theta_{\delta_{}}}\ F\bigr)+
i\ {\bar \theta}{^{\dot \alpha_{}}}\
{\partial_{m_{}}}\bigl({\theta^{\delta_{}}}\ {\theta_{\delta_{}}}\
F\bigr)\ {\sigma^{m_{}}{}_{\gamma_{}}{}_{\dot \alpha_{}}}$
}

\noindent
{\funfn ex = initdis[ ex, oddlist ]}

$-$ ${\theta^{\delta_{}}}\ (-$ ${\theta_{\delta_{}}}\
{\partial_{\gamma_{}}}\bigl(F\bigr))+$
${\partial_{\gamma_{}}}\bigl({\theta^{\delta_{}}}\bigr)\
{\theta_{\delta_{}}}\ F-$ ${\theta^{\delta_{}}}\
{\partial_{\gamma_{}}}\bigl({\theta_{\delta_{}}}\bigr)\ F+$ $i\ {\bar
\theta}{^{\dot \alpha_{}}}\
{\partial_{m_{}}}\bigl({\theta^{\delta_{}}}\bigr)\
{\theta_{\delta_{}}}\ F\ {\sigma^{m_{}}{}_{\gamma_{}}{}_{\dot
\alpha_{}}}+$

\centerline{ $i\ {\bar \theta}{^{\dot \alpha_{}}}\
{\theta^{\delta_{}}}\
{\partial_{m_{}}}\bigl({\theta_{\delta_{}}}\bigr)\ F\
{\sigma^{m_{}}{}_{\gamma_{}}{}_{\dot \alpha_{}}}+$ $i\ {\bar
\theta}{^{\dot \alpha_{}}}\ {\theta^{\delta_{}}}\
{\theta_{\delta_{}}}\ {\partial_{m_{}}}\bigl(F\bigr)\
{\sigma^{m_{}}{}_{\gamma_{}}{}_{\dot \alpha_{}}}$ }

\noindent
{\funfn     ex = oex2lists[ ex ]}

\noindent
{\funfn ex = raise[ \#, {gras, grasd, lor}, raiselist ]\& /@ ex}

$-$ $1,{\theta^{\delta_{}}},-$
$1,{\epsilon_{\delta_{}}{}_{\alpha_{}}},
{\theta^{\alpha_{}}},{\partial_{\gamma_{}}}\bigl(F\bigr)$

${\partial_{\gamma_{}}}\bigl({\theta^{\delta_{}}}\bigr),
{\epsilon_{\delta_{}}{}_{\alpha_{}}},{\theta^{\alpha_{}}},F$

$-$ $1,{\theta^{\delta_{}}},{\epsilon_{\delta_{}}{}_{\alpha_{}}},
{\partial_{\gamma_{}}}\bigl({\theta^{\alpha_{}}}\bigr),F\hfill$

$i,{\bar \theta}{^{\dot \alpha_{}}},{\partial_{m_{}}}\bigl({
\theta^{\delta_{}}}\bigr),{\epsilon_{\delta_{}}{}_{\alpha_{}}},
{\theta^{\alpha_{}}},F,{\sigma^{m_{}}{}_{\gamma_{}}{}_{\dot
\alpha_{}}}$

$i,{\bar \theta}{^{\dot \alpha_{}}},{\theta^{\delta_{}}},
{\epsilon_{\delta_{}}{}_{\alpha_{}}},
{\partial_{m_{}}}\bigl({\theta^{\alpha_{}}}\bigr),
F,{\sigma^{m_{}}{}_{\gamma_{}}{}_{\dot \alpha_{}}}$

$i,{\bar \theta}{^{\dot
\alpha_{}}},{\theta^{\delta_{}}},{\epsilon_{\delta_{}}
{}_{\alpha_{}}},{\theta^{\alpha_{}}},{\partial_{m_{}}}\bigl(F\bigr),
{\sigma^{m_{}}{}_{\gamma_{}}{}_{\dot \alpha_{}}}$

\noindent
{\funfn ex = dman[ ex ]}

$\delta_{\gamma_{}}{}^{\delta_{}},{\epsilon_{\delta_{}}
{}_{\alpha_{}}},{\theta^{\alpha_{}}},F\hfill$

$-$ $1,{\theta^{\delta_{}}},{\epsilon_{\delta_{}}{}_{\alpha_{}}},
\delta_{\gamma_{}}{}^{\alpha_{}},F\hfill$

$i,{\bar \theta}{^{\dot \alpha_{}}},{\theta^{\delta_{}}},
{\epsilon_{\delta_{}}{}_{\alpha_{}}},{\theta^{\alpha_{}}},
{\partial_{m_{}}}\bigl(F\bigr),{\sigma^{m_{}} {}_{\gamma_{}}{}_{\dot
\alpha_{}}}\hfill$

\noindent
{\funfn ex = deltaman[ \#, {lor, gras, grasd} ]\& /@ ex}

${\epsilon_{\gamma_{}}{}_{\alpha_{}}},{\theta^{\alpha_{}}},F\hfill$

$-$
$1,{\theta^{\delta_{}}},{\epsilon_{\delta_{}}{}_{\gamma_{}}},F\hfill$

$i,{\bar \theta}{^{\dot \alpha_{}}},{\theta^{\delta_{}}},
{\epsilon_{\delta_{}}{}_{\alpha_{}}},{\theta^{\alpha_{}}},
{\partial_{m_{}}}\bigl(F\bigr),{\sigma^{m_{}} {}_{\gamma_{}}{}_{\dot
\alpha_{}}}\hfill$

\noindent
{\funfn ex = lists2fulls[ex, funlist, oddlist]}

\noindent
{\funfn ex = canon[\#, out, funlist, oddlist]\& /@ ex}

\noindent
{\funfn ex = zero[ex]}

\noindent
{\funfn ex = fulls2oex[ ex ]}

$-$ $2\ {\theta^{\alpha_{}}}\ {\epsilon_{\alpha_{}}{}_{\gamma_{}}}\
F+$ $i\ {\theta^{\alpha_{}}}\ {\theta^{\beta_{}}}\ {\bar
\theta}{^{\dot \alpha_{}}}\ {\epsilon_{\alpha_{}}{}_{\beta_{}}}\
{\sigma^{m_{}}{}_{\gamma_{}}{}_{\dot \alpha_{}}}\
{\partial_{m_{}}}\bigl(F\bigr)\hfill$

\abvt
\noindent
{\Large\bf
Appendix D: Noninteractive Use of the Package
}
\belt

In this example, the expression:

\centerline{
${\bar D_{\dot \beta_{}}}\Phi$
}

\vskip 10 pt
\noindent
is evaluated using the program {\funfn work} that contains the
functions from the package. The program is shown first, followed by the
output. The arguments of the output commands are set to show the
output on a screen. The output presented here is exactly the same as
the output produced on a screen in the test run. The previously
defined lists are the same as in Appendix~C.

The program is:

\vskip 10 pt
{\funfn
    ex = DSSb[ $\{$betad, d$\}$, Fi[] ];

    out = outopen[ $\{$some, "stdout", tex, tog$\}$ ];

	writework[ out[[2]], "begin:", $\{$ex$\}$ ];

    ex = wbfields[ ex ];

	writework[ out[[2]], "wbfields:", $\{$ex$\}$ ];

    ex = initdis[ ex, oddlist ];

	writework[ out[[1]], "initdis:", $\{$ex$\}$ ];

    ex = oex2lists[ ex ];

	writework[ out[[1]], "oex2lists:", ex ];

    ex = raise[ \#, $\{$gras, grasd, lor$\}$, raiselist ]\& /@ ex;

	writework[ out[[1]], "up:", ex];

    ex = dman[ ex ];

	writework[ out[[1]], "dman:", ex ];

    ex = metricman[ \#, $\{$lor, gras, grasd$\}$ ]\& /@ ex;

	writework[ out[[1]], "metricman:", ex ];

    ex = deltaman[ \#, $\{$lor, gras, grasd$\}$ ]\& /@ ex;

	writework[ out[[2]], "deltaman:", ex ];

    ex = lists2fulls[ex, funlist, oddlist];

	writework[ out[[1]], "lists2fulls:", ex];

    ex = toodd[ \#, gras, oddlist ]\& /@ ex;

    ex = toodd[ \#, grasd, oddlist ]\& /@ ex;

    ex = zero[ex];

	writework[ out[[2]], "zero:", ex ];

    ex = ssigma[ ex, funlist, oddlist ];

	writework[ out[[2]],  "ssigma:", ex ];

    ex = canon[\#, out, funlist, oddlist]\& /@ ex;

	writework[ out[[1]], "canon:", ex];

    ex = zero[ex];

	writework[ out[[1]], "zero:", ex ];

    ex = fulls2oex[ ex ];

	writework[ out[[3]],  "end:", $\{$ex$\}$ ];

    outclose[ $\{$some, "stdout", tex, tog$\}$ ];
}

\vskip 10 pt
The output is:

\vskip 10 pt
\noindent$\hbox{begin:}\hfill$

${\bar D_{\dot \beta_{}}}\bigl(\Phi\bigr)\hfill$

\noindent$\hbox{wbfields:}\hfill$

\noindent
\centerline{ $-$ ${\partial_{\dot \beta_{}}}\bigl(A+$ $2^{1/2}\
{\theta^{\alpha_{}}}\ {\psi_{\alpha_{}}}+$ $F\ {\theta^{\alpha_{}}}\
{\theta_{\alpha_{}}}+$ $i\ {\partial_{m_{}}}\bigl(A\bigr)\
{\theta^{\alpha_{}}}\ {\sigma^{m_{}}{}_{\alpha_{}}{}_{\dot
\alpha_{}}}\ {\bar \theta}{^{\dot \alpha_{}}}+$
$({\partial_{m_{}}}\bigl({\partial_{n_{}}}\bigl(A\bigr)\bigr)\
{\eta^{m_{}}{}^{n_{}}}\ {\theta^{\alpha_{}}}\ {\theta_{\alpha_{}}}\
{\bar \theta}{_{\dot \alpha_{}}}\ {\bar \theta}{^{\dot
\alpha_{}}})/4-$ }

\noindent
\centerline{ $(i\ {\theta^{\alpha_{}}}\ {\theta_{\alpha_{}}}\
{\partial_{m_{}}}\bigl({\psi^{\beta_{}}}\bigr)\ {\bar \theta}{^{\dot
\gamma_{}}}\ {\sigma^{m_{}}{}_{\beta_{}}{}_{\dot
\gamma_{}}})/2^{1/2}\bigr)-$ $i\ ({\sigma^{k_{}}{}_{\gamma_{}}{}_{\dot
\beta_{}}}\ {\theta^{\gamma_{}}})\ {\partial_{k_{}}}\bigl(A+$
$2^{1/2}\ {\theta^{\alpha_{}}}\ {\psi_{\alpha_{}}}+$ $F\
{\theta^{\alpha_{}}}\ {\theta_{\alpha_{}}}+$ }

\noindent
\centerline{ $i\ {\partial_{m_{}}}\bigl(A\bigr)\ {\theta^{\alpha_{}}}\
{\sigma^{m_{}}{}_{\alpha_{}}{}_{\dot \alpha_{}}}\ {\bar \theta}{^{\dot
\alpha_{}}}+$
$({\partial_{m_{}}}\bigl({\partial_{n_{}}}\bigl(A\bigr)\bigr)\
{\eta^{m_{}}{}^{n_{}}}\ {\theta^{\alpha_{}}}\ {\theta_{\alpha_{}}}\
{\bar \theta}{_{\dot \alpha_{}}}\ {\bar \theta}{^{\dot
\alpha_{}}})/4-$ $(i\ {\theta^{\alpha_{}}}\ {\theta_{\alpha_{}}}\
{\partial_{m_{}}}\bigl({\psi^{\beta_{}}}\bigr)\ {\bar \theta}{^{\dot
\gamma_{}}}\ {\sigma^{m_{}}{}_{\beta_{}}{}_{\dot
\gamma_{}}})/2^{1/2}\bigr)$ }

\noindent$\hbox{deltaman:}\hfill$

$1/4,{\partial_{m_{}}}\bigl({\partial_{n_{}}}\bigl(A\bigr)\bigr),
{\eta^{m_{}}{}^{n_{}}},{\theta^{\alpha_{}}},-$
$1,{\epsilon_{\alpha_{}}{}_{\beta_{}}},{\theta^{\beta_{}}},-$
$1,{\epsilon_{\dot \beta_{}}{}_{\dot \gamma_{}}},{\bar \theta}{^{\dot
\gamma_{}}}\hfill$

$1/4,{\partial_{m_{}}}\bigl({\partial_{n_{}}}\bigl(A\bigr)\bigr),
{\eta^{m_{}}{}^{n_{}}},{\theta^{\alpha_{}}},-$
$1,{\epsilon_{\alpha_{}}{}_{\beta_{}}},{\theta^{\beta_{}}},
{\epsilon_{\dot \alpha_{}}{}_{\dot \beta_{}}},{\bar \theta}{^{\dot
\alpha_{}}}\hfill$

$-$ $i,{\sigma^{k_{}}{}_{\gamma_{}}{}_{\dot \beta_{}}},
{\theta^{\gamma_{}}},{\partial_{k_{}}}\bigl(A\bigr)\hfill$

$-$ $i,{\sigma^{k_{}}{}_{\gamma_{}}{}_{\dot
\beta_{}}},{\theta^{\gamma_{}}}, 2^{1/2},{\theta^{\alpha_{}}},
{\epsilon_{\alpha_{}}{}_{\beta_{}}},
{\partial_{k_{}}}\bigl({\psi^{\beta_{}}}\bigr)\hfill$

$-$ $i,{\sigma^{k_{}}{}_{\gamma_{}}{}_{\dot \beta_{}}},
{\theta^{\gamma_{}}},{\partial_{k_{}}}\bigl(F\bigr),
{\theta^{\alpha_{}}},{\epsilon_{\alpha_{}}{}_{\beta_{}}},
{\theta^{\beta_{}}}\hfill$

$-$ $i,{\sigma^{k_{}}{}_{\gamma_{}}{}_{\dot
\beta_{}}},{\theta^{\gamma_{}}},i,
{\partial_{k_{}}}\bigl({\partial_{m_{}}}\bigl(A\bigr)\bigr),
{\theta^{\alpha_{}}},{\sigma^{m_{}}{}_{\alpha_{}} {}_{\dot
\alpha_{}}},{\bar \theta}{^{\dot \alpha_{}}}\hfill$

$-$ $i,{\sigma^{k_{}}{}_{\gamma_{}}{}_{\dot
\beta_{}}},{\theta^{\gamma_{}}},1/4,{\partial_{k_{}}}
\bigl({\partial_{m_{}}}\bigl({\partial_{n_{}}}
\bigl(A\bigr)\bigr)\bigr),{\eta^{m_{}}{}^{n_{}}},
{\theta^{\alpha_{}}},{\epsilon_{\alpha_{}}{}_{\beta_{}}},
{\theta^{\beta_{}}},{\epsilon_{\dot \alpha_{}}{}_{\dot
\gamma_{}}},{\bar \theta}{^{\dot \gamma_{}}},{\bar \theta}{^{\dot
\alpha_{}}}\hfill$

$-$ $i,{\sigma^{k_{}}{}_{\gamma_{}}{}_{\dot
\beta_{}}},{\theta^{\gamma_{}}},-$ $i,2^(-$
$1/2),{\theta^{\alpha_{}}},{\epsilon_{\alpha_{}}
{}_{\delta_{}}},{\theta^{\delta_{}}},
{\partial_{k_{}}}\bigl({\partial_{m_{}}}
\bigl({\psi^{\beta_{}}}\bigr)\bigr), {\bar \theta}{^{\dot
\gamma_{}}},{\sigma^{m_{}}{}_{\beta_{}}{}_{\dot \gamma_{}}}\hfill$

$i,{\partial_{m_{}}}\bigl(A\bigr),{\theta^{\alpha_{}}},{\sigma^{m_{}}
{}_{\alpha_{}}{}_{\dot \beta_{}}}\hfill$

$-$ $i,2^(-$ $1/2),{\theta^{\alpha_{}}},-$
$1,{\epsilon_{\alpha_{}}{}_{\gamma_{}}},{\theta^{\gamma_{}}},-$
$1,{\partial_{m_{}}}\bigl({\psi^{\beta_{}}}\bigr),{\sigma^{m_{}}
{}_{\beta_{}}{}_{\dot \beta_{}}}\hfill$

\noindent$\hbox{zero:}\hfill$

$1,1/4,-$ $1,-$ $1,{\theta^{\alpha_{}}},{\theta^{\beta_{}}},{\bar
\theta} {^{\dot
\gamma_{}}},{\eta^{m_{}}{}^{n_{}}},{\epsilon_{\alpha_{}}
{}_{\beta_{}}},{\epsilon_{\dot \beta_{}}{}_{\dot \gamma_{}}},
{\partial_{m_{}}}\bigl({\partial_{n_{}}}\bigl(A\bigr)\bigr)\hfill$

$1,1/4,-$ $1,{\theta^{\alpha_{}}},{\theta^{\beta_{}}},{\bar
\theta}{^{\dot
\alpha_{}}},{\eta^{m_{}}{}^{n_{}}},{\epsilon_{\alpha_{}}
{}_{\beta_{}}},{\epsilon_{\dot \alpha_{}}{}_{\dot
\beta_{}}},{\partial_{m_{}}}\bigl({\partial_{n_{}}}
\bigl(A\bigr)\bigr)\hfill$

$1,-$ $i,{\theta^{\gamma_{}}},{\sigma^{k_{}}{}_{\gamma_{}}{}_{\dot
\beta_{}}},{\partial_{k_{}}}\bigl(A\bigr)\hfill$

$1,-$ $i,2^{1/2},{\theta^{\gamma_{}}},{\theta^{\alpha_{}}},
{\epsilon_{\alpha_{}}{}_{\beta_{}}},{\sigma^{k_{}}{}_{\gamma_{}}{}_{\dot
\beta_{}}},{\partial_{k_{}}}\bigl({\psi^{\beta_{}}}\bigr)\hfill$

$1,-$ $i,i,{\theta^{\gamma_{}}},{\theta^{\alpha_{}}},{\bar
\theta}{^{\dot \alpha_{}}},{\sigma^{k_{}}{}_{\gamma_{}}{}_{\dot
\beta_{}}},{\sigma^{m_{}}{}_{\alpha_{}}{}_{\dot
\alpha_{}}},{\partial_{k_{}}}\bigl({\partial_{m_{}}}
\bigl(A\bigr)\bigr)\hfill$

$1,i,{\theta^{\alpha_{}}},{\sigma^{m_{}}{}_{\alpha_{}} {}_{\dot
\beta_{}}},{\partial_{m_{}}}\bigl(A\bigr)\hfill$

$1,-$ $i,2^(-$ $1/2),-$ $1,-$
$1,{\theta^{\alpha_{}}},{\theta^{\gamma_{}}},{\epsilon_{\alpha_{}}
{}_{\gamma_{}}},{\sigma^{m_{}}{}_{\beta_{}}{}_{\dot
\beta_{}}},{\partial_{m_{}}}\bigl({\psi^{\beta_{}}}\bigr)\hfill$

\noindent$\hbox{ssig:}\hfill$

$1,1,1/4,-$ $1,-$ $1,{\theta^{\alpha_{}}},{\theta^{\beta_{}}},{\bar
\theta}{^{\dot
\gamma_{}}},{\eta^{m_{}}{}^{n_{}}},{\epsilon_{\alpha_{}}
{}_{\beta_{}}},{\epsilon_{\dot \beta_{}}{}_{\dot
\gamma_{}}},{\partial_{m_{}}}\bigl({\partial_{n_{}}}
\bigl(A\bigr)\bigr)\hfill$

$1,1,1/4,-$ $1,{\theta^{\alpha_{}}},{\theta^{\beta_{}}},{\bar
\theta}{^{\dot
\alpha_{}}},{\eta^{m_{}}{}^{n_{}}},{\epsilon_{\alpha_{}}
{}_{\beta_{}}},{\epsilon_{\dot \alpha_{}}{}_{\dot
\beta_{}}},{\partial_{m_{}}}\bigl({\partial_{n_{}}}
\bigl(A\bigr)\bigr)\hfill$

$1,1,-$ $i,{\theta^{\gamma_{}}},{\sigma^{k_{}}{}_{\gamma_{}} {}_{\dot
\beta_{}}},{\partial_{k_{}}}\bigl(A\bigr)\hfill$

$1,1,-$ $i,2^{1/2},{\theta^{\gamma_{}}},{\theta^{\alpha_{}}},
{\epsilon_{\alpha_{}}{}_{\beta_{}}},{\sigma^{k_{}}
{}_{\gamma_{}}{}_{\dot \beta_{}}},{\partial_{k_{}}}
\bigl({\psi^{\beta_{}}}\bigr)\hfill$

$1,1,-$ $i,i,{\theta^{\gamma_{}}},{\theta^{\alpha_{}}},{\bar
\theta}{^{\dot
\alpha_{}}},{\sigma_{+++}{}^{k_{}}{}^{m_{}}{}_{\gamma_{}}
{}_{\alpha_{}}{}_{\dot \beta_{}}{}_{\dot \alpha_{}}},{\partial_{k_{}}}
\bigl({\partial_{m_{}}}\bigl(A\bigr)\bigr)\hfill$

$1,1,-$ $i,i,{\theta^{\gamma_{}}},{\theta^{\alpha_{}}},{\bar
\theta}{^{\dot
\alpha_{}}},{\sigma_{-+-}{}^{k_{}}{}^{m_{}}{}_{\gamma_{}}
{}_{\alpha_{}}{}_{\dot \beta_{}}{}_{\dot
\alpha_{}}},{\partial_{k_{}}}\bigl({\partial_{m_{}}}
\bigl(A\bigr)\bigr)\hfill$

$1,1,-$ $i,i,{\theta^{\gamma_{}}},{\theta^{\alpha_{}}},{\bar
\theta}{^{\dot \alpha_{}}},{\sigma_{--+}{}^{k_{}}{}^{m_{}}
{}_{\gamma_{}}{}_{\alpha_{}}{}_{\dot \beta_{}}{}_{\dot
\alpha_{}}},{\partial_{k_{}}}\bigl({\partial_{m_{}}}
\bigl(A\bigr)\bigr)\hfill$

$1,1,-$ $i,i,-$ $1/2,{\theta^{\gamma_{}}},{\theta^{\alpha_{}}},{\bar
\theta}{^{\dot
\alpha_{}}},{\eta^{k_{}}{}^{m_{}}},{\epsilon_{\gamma_{}}
{}_{\alpha_{}}},{\epsilon_{\dot \beta_{}}{}_{\dot
\alpha_{}}},{\partial_{k_{}}}\bigl({\partial_{m_{}}}
\bigl(A\bigr)\bigr)\hfill$

$1,1,i,{\theta^{\alpha_{}}},{\sigma^{m_{}}{}_{\alpha_{}} {}_{\dot
\beta_{}}},{\partial_{m_{}}}\bigl(A\bigr)\hfill$

$1,1,-$ $i,2^(-$ $1/2),-$ $1,-$
$1,{\theta^{\alpha_{}}},{\theta^{\gamma_{}}},{\epsilon_{\alpha_{}}
{}_{\gamma_{}}},{\sigma^{m_{}}{}_{\beta_{}}{}_{\dot
\beta_{}}},{\partial_{m_{}}}\bigl({\psi^{\beta_{}}}\bigr)\hfill$

\noindent$\hbox{end:}\hfill$

$(-$ $i\ {\theta^{\alpha_{}}}\ {\theta^{\beta_{}}}\
{\epsilon_{\alpha_{}}{}_{\beta_{}}}\
{\sigma^{m_{}}{}_{\gamma_{}}{}_{\dot \beta_{}}}\
{\partial_{m_{}}}\bigl({\psi^{\gamma_{}}}\bigr))/2^{1/2}+$ $i\
2^{1/2}\ {\theta^{\alpha_{}}}\ {\theta^{\beta_{}}}\
{\epsilon_{\alpha_{}}{}_{\gamma_{}}}\
{\sigma^{m_{}}{}_{\beta_{}}{}_{\dot \beta_{}}}\
{\partial_{m_{}}}\bigl({\psi^{\gamma_{}}}\bigr)\hfill$

\vskip 10 pt
Due to a property of the $4-D,\ N=1$ model, the last expression equals
zero. However, since the package in its current form does not contain
a function that would use the property, the initial expression is not
simplified to the fullest extent. If such a function is added to the
program, the initial expression would be calculated to be zero.

\vfill \eject

\end{document}